# Gate-tuned interlayer coupling in van der Waals ferromagnet $Fe_3GeTe_2$ nanoflakes

Guolin Zheng[1#], Wen-Qiang Xie[2#], Sultan Albarakati[1#], Meri Algarni[1], Cheng Tan[1], Yihao Wang[3], Jingyang Peng[1], James Partridge[1], Lawrence Farrar[1], Jiabao Yi[4], Yimin Xiong[3], Mingliang Tian[3,5,6], Yu-Jun Zhao[2*], Lan Wang[1*]

[1]School of Science, RMIT University, Melbourne, VIC 3001, Australia.

[2]Department of Physics, South China University of Technology, Guangzhou 510640, China.

[3]Anhui Province Key Laboratory of Condensed Matter Physics at Extreme Conditions, High Magnetic Field Laboratory, Chinese Academy of Sciences (CAS), Hefei 230031, Anhui, China.

[4]Global Innovative Center for Advanced Nanomaterials, School of Engineering, University of Newcastle, Callaghan, NSW 2308, Australia.

[5]Department of Physics, School of Physics and Materials Science, Anhui University, Hefei 230601, Anhui, China.

[6]Collaborative Innovation Center of Advanced Microstructures, Nanjing University, Nanjing 210093, China.

[#] Those authors equally contribute to the paper.

[*] Corresponding authors. Correspondence and requests for materials should be addressed to Y.-J. Z. (email: zhaoyj@scut.edu.cn) and L. W. (email: lan.wang@rmit.edu.au).

# Abstract

The weak interlayer coupling in van der Waals (vdW) magnets has confined their application to two dimensional (2D) spintronic devices. Here, we demonstrate that the interlayer coupling in a vdW magnet $Fe_3GeTe_2$ (FGT) can be largely modulated by a protonic gate. With the increase of the protons intercalated among vdW layers, interlayer magnetic coupling increases. Due to the existence of antiferromagnetic layers in FGT nanoflakes, the increasing interlayer magnetic coupling induces exchange bias in protonated FGT nanoflakes. Most strikingly, a rarely seen zero-field cooled (ZFC) exchange bias with very large values (maximally up to 1.2 kOe) has been observed when higher positive voltages ($V_g \geq 4.36$ V) are applied to the protonic gate, which clearly demonstrates that a strong interlayer coupling is realized by proton intercalation. Such strong interlayer coupling will enable a wider range of applications for vdW magnets. .

The weak interlayer coupling in vdW materials readily enables exfoliation to explore the 2D limit and stacking for heterostructure devices. With improved techniques for assembling vdW heterostructures and increased availability of high-quality vdW materials, especially vdW magnets [1-4], various new phenomena and devices are rapidly emerging [5-17]. However, the same weak interlayer coupling that provides ease of fabrication limits the application of vdW materials. Strong magnetic coupling in vdW magnets and heterostructures is essential for many novel functional devices. For example, strong magnetic proximity effects observed between ferromagnetic insulators and topological insulators (TI) within heterostructures are a promising method to realize the quantum anomalous Hall effect [18] at high temperature. Similarly, strong coupling between 2D ferromagnetic (FM) and ferroelectric (FE) layers is essential to realize 2D multiferroics [19]. Unfortunately, little progress has been achieved in this important and challenging research direction so far. This is because magnetic exchange coupling requires electron exchange between atoms. However, in magnetic vdW heterostructures, electron exchange is not expected, only dipole-dipole interaction exists. The absence of magnetic exchange coupling, such as the exchange bias effect, in magnetic vdW heterostructures is therefore unsurprising.

The exchange bias effect is one of the most important magnetic coupling effects and has been studied for more than 60 years. The exchange bias effect has been widely observed in metal-based magnetic heterostructures with FM-antiferromagnetic (AFM) [20], FM-ferrimagnetic [21], AFM-ferrimagnetic [22], and FM-spin glass [23] interfaces. It has also been seen in bulk polycrystalline alloy [24], metal thin film heterostructures grown by magnetron sputtering or MBE [25-28] and topological insulator-antiferromagnet heterostructures [29] among others.

In this letter, using exchange bias as a tool, we show that the interlayer coupling in vdW magnet $Fe_3GeTe_2$ nanoflakes can be modulated significantly by a protonic gate. With increasing gate voltage, the interlayer magnetic coupling increases and induces both field-cooled (FC) and a rarely seen ZFC exchange bias effect in the protonated FGT nanoflakes. Our finding suggests the possibility for new and wider vdW heterostructure devices with strong interlayer coupling, an important step towards vdW spintronic applications.

Figure 1A presents the schematic diagram of an electrically gated device placed on a solid proton electrolyte (See Supplementary information). Fig. 1B shows the magnetic field dependent Hall resistance $R_{xy}$ at various gate voltages. As the voltage was swept from -5.5 to 4.4 V, the coercivity of $Fe_3GeTe_2$ nanoflake decreased, while the anomalous Hall resistance $R_{xy}$ increased until the gate voltage reached -4.16 V and then decreased monotonically. $R_{xy}$ is proportional to magnetization so the change in $R_{xy}$ shows the variation in magnetization with applied gate voltage. As reported previously [30], the nearly square-shaped magnetic loop of FGT is a characteristic of a large perpendicular anisotropy. A clear exchange bias effect emerges at $V_g = 4.4$ V, which will be discussed later. The temperature dependence of the magnetic phase of the pristine FGT flake is shown in Fig 1C. With a 1 T magnetic field we first saturated the $R_{xy}$. The field was then reduced to zero and the temperature dependent $R_{xy}$ was examined. As shown in Fig. 1C, $R_{xy}$ decreases as the temperature decreases from 60 K and 20 K (shown surrounded by the black dashed circle), in contrast to other temperature regions. This previously observed anomalous behaviour [4, 30] indicates the coexistence of antiferromagnetic (AFM) and ferromagnetic (FM) phases in FGT [31]. These coexistent phases should generate the exchange bias effect after field cooling. The lack of exchange bias in the pristine FGT nanoflakes reveals a weak AFM-FM coupling.

When the proton concentrations in FGT nanoflakes are increased by increasing the voltage to 4.3 V, the exchange bias effect emerges, as shown in Fig. 2A. At $V_g = 4.3$ V, the sample was cooled from 200 to 2 K with applied fields of ± 1 T. The magnetic hysteresis loop was found to shift to negative and positive fields, respectively, consistent with standard negative exchange bias behaviour. The 'training effect', where the exchange bias decreased or disappeared after several measurements, was also observed (see Fig. 2B). The observation of the exchange bias effect in the protonated FGT nanoflakes indicates a dramatic enhancement of interlayer AFM-FM coupling.

Further unexpected magnetic hysteresis behaviour was observed with higher applied gate voltage. At $V_g$ = 4.36 V, the exchange bias effect was realized for both zero-field cooling and ±1 T field cooling, as shown in Fig. 3A-3C. After the initially zero-field cooling to 2 K, the magnetic hysteresis loop measurements were repeated 16 times. Both positive and negative exchange bias effects were randomly displayed and all the hysteresis loops were approximately square-shaped, excepting a hysteresis tail which featured in several loops (indicated by the black arrows). Although the 16 loops were measured continuously, the standard training effect was not observed. Instead, the 16 nearly square-shaped loops displayed large and correlated values of coercivity and exchange bias. Fig.3B and Fig. 3C show the evolution of the coercivity and exchange-bias amplitude with the measurement times for zero-field cooling and ± 1 T cooling, respectively. As shown, the cooling process is unimportant at this gate voltage. With the exception of some loops with large coercivities but small exchange bias amplitude or vice versa (shown by red dashed circles in the figure), the coercivity and bias amplitude in the $Fe_3GeTe_2$ were correlated, namely, large exchange bias induced large coercivity with exchange bias up to ~ 1 kOe. The same correlation between coercivity and exchange bias was also observed at $V_g$ = 4.4 V (Supplementary fig. S6). The large exchange bias clearly demonstrates that large interlayer coupling is realized through protonic gating.

To observe the exchange bias effect, unidirectional anisotropy must be induced by AFM/FM interaction at the interface. ZFC exchange bias was first observed in an InMnNi alloy system previously [24]. The ferromagnetic domains in InMnNi alloy system expand and then couple with each other under an applied magnetic field. In this process, a new AFM/FM interface is formed and unidirectional anisotropy is induced. In Fig. 3A, large ZFC exchange bias was observed in a protonated FGT nanoflake, suggesting that a new AFM/FM interface was formed as the magnetic field was increased. However, unlike the InMnNi system, the protonated FGT flake showed nearly square-shaped loops with similar saturated $R_{xy}$ in all measurements, regardless of the differing coercivities i.e. the saturation magnetization of FGT is constant. Hence, the formation of a new FM/AFM interface occurs by a mechanism alternative to that in the alloyed InMnNi system. We hypothesize, the ZFC exchange bias is induced if the AFM and FM coupling energies are similar. A large positive gate voltage induces a proton concentration gradient along the interlayer direction and the energies of FM and AFM coupling at certain interfaces are approximately equal. Under these conditions unidirectional anisotropy can occur between the FM and AFM domains. As shown in the schematic diagrams in Fig. 3D, FGT layers with nearly identical AFM and FM coupling energy support the existence of both FM domains and AFM domains. At the AFM/FM interface, coupling can be transferred between FM and AFM due to the small energy discrepancy. This transfer is always accompanied by a new AFM/FM interface. For example, a negative exchange bias can be formed if the AFM/FM interface coupling is of the FM type. While exchange bias will be positive if the AFM/FM interface is AFM. It is also possible that the coupling at the interface changes from FM to AFM or vice-versa in a single field-sweeping loop to produce various coercivities and exchange bias values. For example, large coercivities with very small exchange bias (or vice versa) are indicated by the red dashed circles in Fig. 3B and Fig. 3C.

The emergence of both FC and ZFC exchange bias in protonated FGT is consistent with the calculations based on density functional theory [32]. To find the energetically favoured configuration of FGT with intercalated H atoms, we have simulated seven possible intercalating sites, as illustrated in Figure S8. The corresponding formation enthalpies (ΔH) with and without inclusion of the vdW interaction are listed in Table S1. These calculated formation enthalpies are all positive, implying that the intercalation of H may require additional assistance (such as gate voltage) for stabilization. The positive enthalpy reveals that the coupling strength at the H-enhanced interface falls below that of the H-H interaction in a hydrogen molecule, which was adopted as a reference in the calculation of enthalpy. Nevertheless, the most stable configuration is adopted for the following calculations, as illustrated in Figure 4. Interestingly, the simulations show that the intercalated H gains electrons from FGT (mainly transferred from Te ions) and is stabilized in the system as $H^-$ (based on Bader charge analysis).

Simulations using a bulk model support the assertion that the intercalated $H^-$ significantly enhances the interaction among the FGT layers (see Fig. S10). When the vdW effect is considered in the calculation, the coupling energy of FGT increases from 0.99 eV to 1.07 eV when a single $H^-$ is intercalated in each unit cell of FGT (See Table S3). Even if the vdW effect is neglected, the coupling energy increases from 0.107 eV to 0.124 eV after a single $H^-$ is intercalated in each unit cell. The enhanced coupling is further confirmed by the double-layer slab model, which directly simulates the coupling of two layers of FGT with intercalated $H^-$ (as discussed in the supplementary). Meanwhile, the distance between the Te layers is also significantly reduced when $H^-$ is intercalated. Considering the vdW effect the reduction is from 8.02 to 7.78 Å and without considering vdW interaction it is 8.57 to 8.40 Å, as illustrated in Figure 4.

Regarding the magnetic coupling among the FGT layers, it turns out that when considering the vdW effect, both FGT and FGT:$H^-$ slightly favour the AFM coupling. Interestingly, when the vdW interaction is neglected, the FGT and FGT:$H^-$ systems are slightly in favour of FM coupling. The results (Table S4 and Figure S12) are further discussed in the supporting information. Band structure calculations describe ideal conditions and cannot fully agree with experiments. However, these calculation results demonstrate two key points. Firstly, the energy discrepancy between AFM and FM coupling in both FGT and FGT:H$^-$ is small, which explains the coexisting of FM and AFM in FGT nanoflakes. Secondly, when the vdW effect is considered, AFM in FGT is more energetically favourable. Without the proton intercalation, the vdW coupling is weak and FGT nanoflakes behave like a FM with limited AFM coupling. With increasing proton intercalation, the vdW coupling becomes stronger and therefore induces more AFM couplings. This results in the decrease of saturation magnetization shown in Fig. 1B. When the proton concentration increases to a certain level, the coupling between AFM and FM layers is strong enough to generate exchange bias as shown in Fig. 2. Further increase of proton concentration will finally induce very strong AFM/FM, FM/FM, AFM/AFM coupling among layers with larger exchange bias. Due to the small energy discrepancy between AFM and FM coupling at the interfaces between AFM and FM domains, the interface coupling may transfer between the AFM and FM and hence ZFC exchange bias is realized.

Using exchange bias as a tool, we have realized for the first time strong interlayer coupling by proton intercalation in vdW magnet FGT nanoflakes. Our finding establishes that proton intercalation is a promising method for much wider vdW heterostructure devices with strong interface coupling and better device performance, such as 2D FM insulator-TI heterostructures for high temperature quantum anomalous Hall effect, and FM-FE heterostructures for 2D multiferroics.

## Acknowledgments

Work at RMIT university was supported by the Australia Research Council Centre of Excellence in Future Low-Energy Electronics Technologies (CE170100039). Work at SCUT was supported by NSFC (GrantNos.11574088 and 51621001).

**Figure captions**

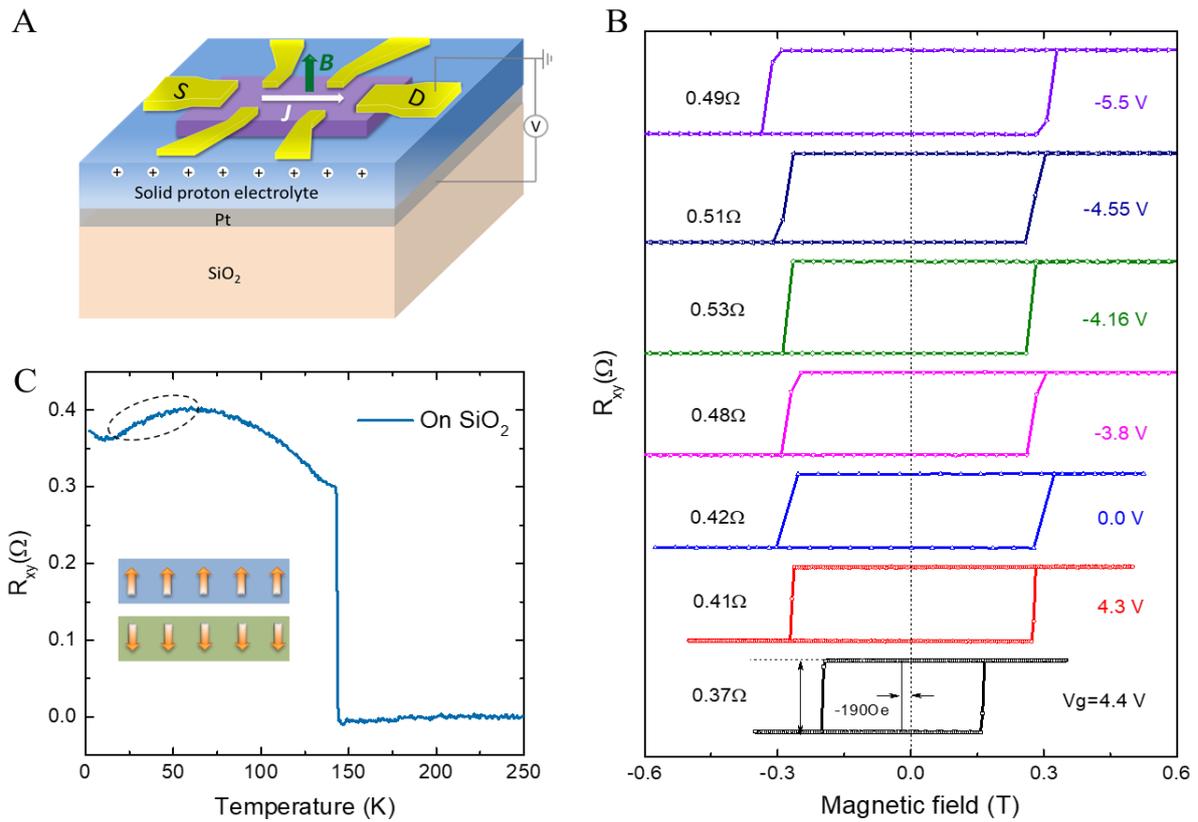

FIG. 1 (A) Schematic of the Hall-bar device on solid proton conductor used for measurements in (B) and (C) in which the current density is *J* and a perpendicular magnetic field *B* is applied. (B) Gate-tuned ferromagnetism in FGT nanoflake (thickness 115 nm, determined by Atomic Force Microscopy). A positive gate voltage decreases both magnetization and coercivity. (C) Remanent Hall resistance $R_{xy}$ as a function of temperature. Inset: Schematic of possible antiferromagnetic phase in pristine FGT.

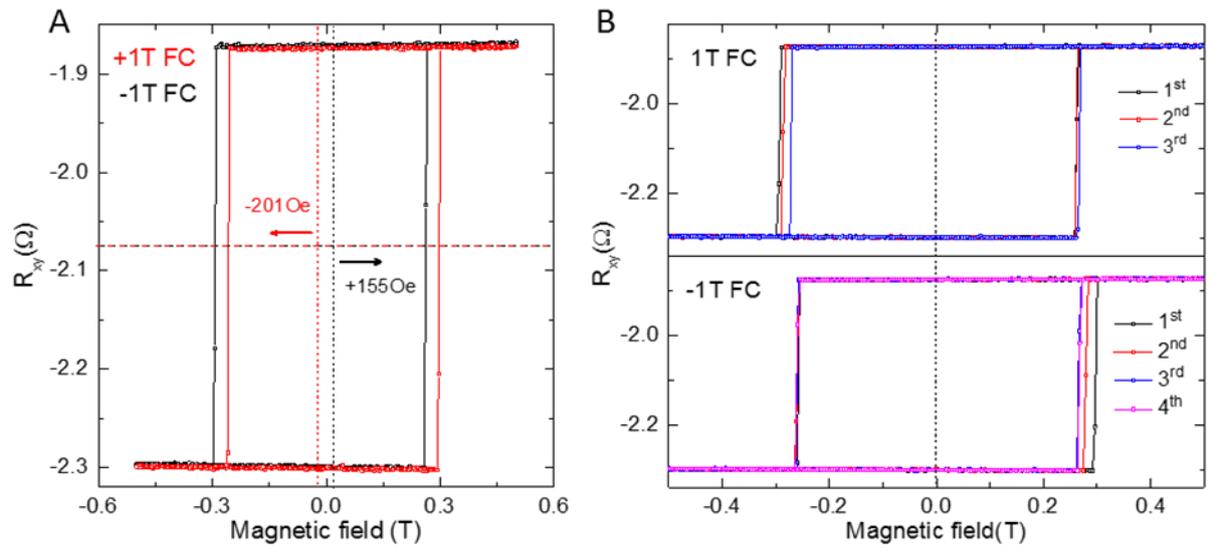

FIG.2 (A) Exchange bias effect after ±1 T field cooling. (B) Training effect of field-cooled exchange bias. Exchange disappears after several measurements.

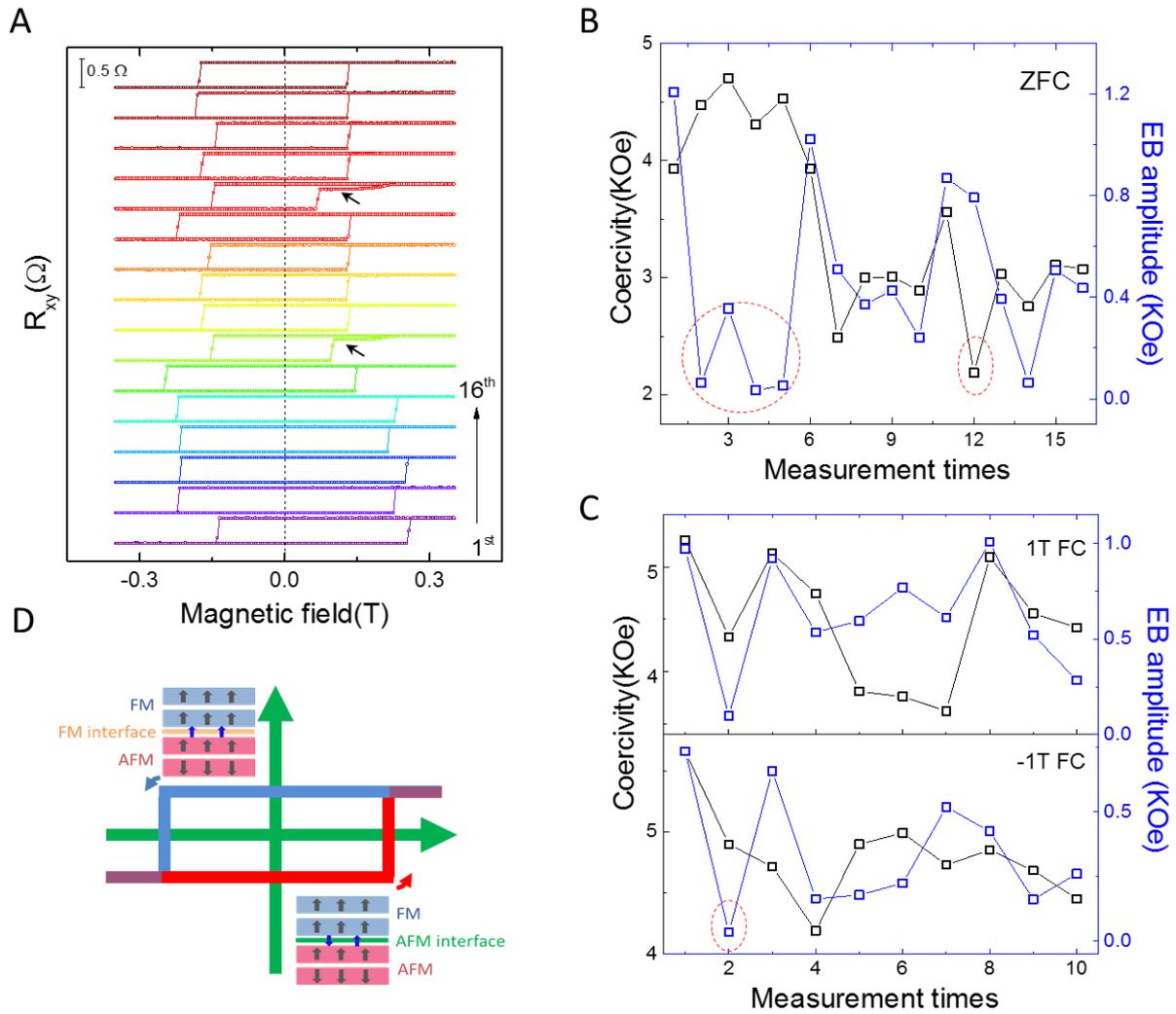

FIG.3 (A) Exchange bias effect after zero-field cooling. Both coercivity and exchange bias exhibits random values during different measurement loops at 2 K. (B) The correlation between coercivity and exchange-bias (EB) amplitude after zero-field cooling. (C) The correlation between coercivity and exchange-bias (EB) amplitude after ±1 T cooling, respectively. (D) Schematic of differing AFM/FM interfaces. If the exchange coupling at the interface changes between FM and AFM in a single field-sweeping loop, unusual magnetic loops with large coercivity but very small exchange bias are observed.

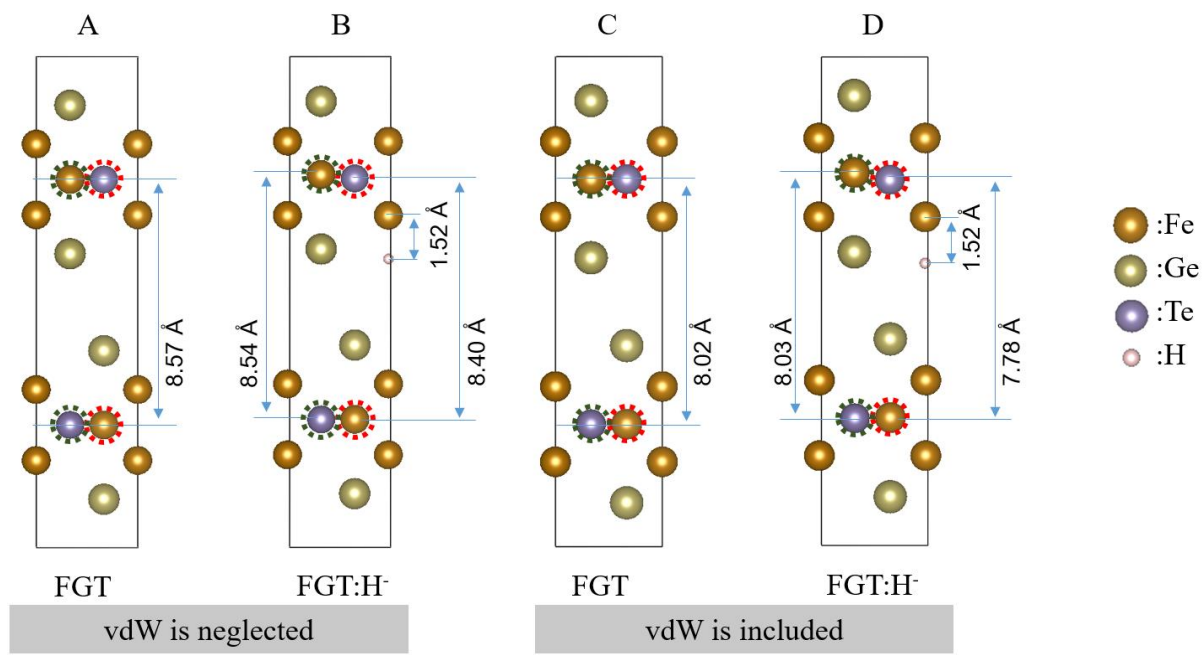

FIG 4. Relaxed structures of (A) FGT and (B) FGT:H⁻ without consideration of the vdW effect. (C) and (D) show similar relaxed structures simulated with the vdW effect included. Upper layer Fe and lower layer Te atoms are highlighted by dotted green (or red) circles to assist in distinguishing the changes in layer separation due to intercalation of H⁻.